\documentclass[twocolumn,showpacs,amsmath,amssymb,letterpaper,jcp]{revtex4}

\usepackage{graphicx}
\usepackage{ctable}
\usepackage{siunitx}

\def\erfc{\mathop{\rm erfc}\nolimits}

\newcommand{\dsurd}[2]{\frac{\partial #1}{\partial #2}}
\newcommand{\sdsurd}[2]{\partial #1/\partial #2}
\newcommand{\ddesurd}[2]{\frac{\partial^2 #1}{{\partial #2}^2}}

\newcommand{\phdeuxo}{p_\text{H$_2$O}}

\newcommand{\psatliq}{p_\text{sat,liq}}
\newcommand{\pspin}{p_\text{spinodal}}
\newcommand{\psat}{p_\text{sat}}
\newcommand{\pspinyukawa}{\pspin^\text{Yukawa}}
\newcommand{\psatyukawa}{\psat^\text{Yukawa}}

\newcommand{\tlinfty}{T_L^\infty}

\newcommand{\gclust}{{g}}
\newcommand{\net}{\gclust^*}

\newcommand{\wet}{W^*}
\newcommand{\ratt}{\beta}
\newcommand{\csd}{f}
\newcommand{\csdeq}{f_\text{eq}}
\newcommand{\cnucl}{F}
\newcommand{\Nclust}{N}
\newcommand{\Nclustquasi}{\Nclust^\text{QS}}
\newcommand{\nquasi}{n^\text{QS}}
\newcommand{\Nclusttransient}{\Nclust^\text{TRANS}}
\newcommand{\ntransient}{n^\text{TRANS}}
\newcommand{\xquasi}{x^\text{QS}}
\newcommand{\xtransient}{x^\text{TRANS}}
\newcommand{\nbclust}{{\mathcal N}}

\newcommand{\puisdix}[2]{#1 \times 10^{#2}}
\newcommand{\tinit}{t_0}
\newcommand{\tend}{t_F}

\newcommand{\tlneg}{\tlinfty = -5^\circ\:\text{C}}
\newcommand{\tlamb}{\tlinfty = 20^\circ\:\text{C}}

\newcommand{\rholsat}{\rho_l^\text{sat}}
\newcommand{\rhovsat}{\rho_v^\text{sat}}
\newcommand{\rholspin}{\rho_l^\text{spin}}
\newcommand{\rhovspin}{\rho_v^\text{spin}}

\begin{document}


\title{On the possibility of homogeneous nucleation of water droplets
  and/or ice crystals during the bounces of a cavitation bubble.}

\author{Olivier Louisnard}
\email{louisnar@mines-albi.fr}
\author{Fabienne Espitalier}

\affiliation{%
  RAPSODEE, UMR EMAC-CNRS 5302, %
  Ecole des Mines d'Albi, Universit\'e de Toulouse, %
  81013 Albi, France%
}


\date{\today}

\begin{abstract}
  Acoustic cavitation is known to trigger ice nucleation in
  supercooled water. Several competing and still debatable mechanisms
  have been proposed in the literature and are related to the pressure
  field in the vicinity of the bubble at the end of its
  collapse. Numerical simulations of the bubble dynamics show that
  during the bubble expansions in the bounces following the main
  collapse, the bubble core temperature reaches values far below
  \SI{0}{\degreeCelsius} for time periods of about \SI{500}{ns}. The
  water vapour present in the bubble during these time intervals
  explores the liquid and the solid region of the phase diagrams
  before going back to the vapour region. On the base of approximate
  nucleation kinetics calculations, we examine to what extent liquid
  droplets could nucleate homogenously in the bubble core during these
  excursions. We also discuss the possibility that the nucleated
  clusters reach the bubble wall and trigger ice nucleation in the
  surrounding liquid if the latter is supercooled.
\end{abstract}

\maketitle

%
\section{Introduction}
%
Gas bubbles submitted to a sound field undergo radial oscillations, a
phenomenon known as acoustic cavitation. For enough large amplitude of
the driving field, such bubbles undergo an explosive growth followed
by a rapid collapse. The density energy in the compressed gas at the
end of the collapse is large enough to break chemical bounds and
produce light emission, known as sonoluminescence.

Cavitation bubbles are also known to trigger ice nucleation in
supercooled water. In absence of ultrasound, nucleation is a
stochastic process and can occur over a relatively wide range of
supercooling temperatures.  When subject to even short ultrasound
bursts, supercooled water can freeze at lower supercooling level
\cite{hem67,huntjacksonNature66,bhadra68,gitlin69,Inada2001_I,%
  chow2005,Nakagawa2006,LindingerMettin2007,Saclier2010exp}.  This
ability to control the nucleation temperature has interesting and
promising industrial applications. Despite various mechanisms have
been proposed to explain this effect and several studies report
experimental results, the mechanism of ice nucleation by a radially
oscillating bubble remains unclear.

Two main available theories have been proposed
\cite{hickling65,hickling94,huntjackson66,huntjacksonNature66}, and
share a common feature: both attribute ice nucleation to a shift of
the freezing point in the supercooled liquid phase, because of large
pressure variations at the end of the bubble collapse. A large
majority of species increase their freezing temperature $T_f$ as
pressure increases. Normal ice (ice I) constitutes an exception since
the solid phase is less dense than the liquid, so that a compression
of supercooled water would therefore normally quench freezing.
Hickling \cite{hickling65} argues that the very large pressures
attained (typically several GPa) in the vicinity of the bubble allow
the nucleation of ice V, VI and VII, which contrarily to ice I, have a
positive $d T_f/d p$ slope. Hunt \& Jackson notices that the large
pressure increase at the end of the collapse is followed by a very
high negative transient pressure, which may increase supersaturation
relative to ice I nucleation. Both theories are equally mentioned in
studies of cavitation-enhanced water freezing. Due to the very short
space- and timescales involved, a direct confirmation of one or the
other mechanism appears hardly feasible, despite Hickling's statement
has been favored by calculations of orders of
magnitude~\cite{OhsakaTrinh97}.

In the present work, we examine the possibility of a third mechanism,
which has been overlooked in past studies. The above-mentioned
theories only examine the thermodynamic state of the surrounding
water. However, in the curse of the volume oscillations of the bubble,
water evaporates and condensates at the interphase, so that the bubble
encloses a variable quantity of water vapor. Finite-rate mass
diffusion of water vapor through noncondensable gas
\cite{storeyszeri2000,toegel2000} and non-equilibrium
evaporation/condensation at the bubble wall
\cite{yasui97,colussi99vapor,storeyszeri2000} are known to prevent
water from condensing during the collapse. This water-trapping
mechanism has been recognized to decrease the final collapse
temperature \cite{moss99computed,yasui2001,storeyszeri2000}, which
plays a crucial role in sonoluminescence, and in particular explains
the enhancement of single-bubble sonoluminescence (SBSL) in cold water
\cite{barberwu1994,vazquezputterman2000,storeyszeriargon,yasui2001}.

However, poor attention has been paid to the fate of the bubble vapor
content after the collapse. A first plausible reason for that is that
sonoluminescence and sonochemistry studies are mainly concerned with
the hot state of the bubble interior at the end of the collapse, so
that the subsequent phases of the bubble dynamics are of little
interest in this framework.  A second reason is that a bubble is not
granted to maintain a spherical shape and even to survive after its
collapse in multi-bubble conditions. However, a single bubble in a
levitation experiment is shape-stable for millions of cycles in
single-bubble experiments \cite{barber,brenner2002}, and bubbles
stable for several cycles have been reported in some multi-bubbles
experiments~\cite{mettin2005}. This justifies the study of the water
vapor content after the main collapse. Of special interest is the
re-expanding phase of the bubble after the primary or secondary
collapses. A naive examination of the radius-time curve around the
collapse shows that the re-expansion also occurs on a timescale almost
as fast as the collapse (see for example Fig.~\ref{figRTzoom}a).  One
may therefore expect that, as the collapse yields considerable heating
of the bubble interior, the re-expansion might produce a large cooling
of the bubble content, including water vapor. The goal of this work is
the assess the latter point and to examine its implications on
potential nucleation of liquid water droplets or ice crystals. From an
historical point of view, this is an attempt to link the long known
effect of cavitation on ice nucleation, to more recent theories
modeling water and heat transport in an oscillating spherical bubble.



%
\section{The state of water vapor during bubble bounces}
%

To calculate the water state in the bubble, we use a simplified model
based on thermal and mass diffusion layers popularized by studies on
SBSL \cite{toegel2000, storeyszeri2001}, and recently validated
against more refined models \cite{StrickerProsperettiLohse2011}. In
order to simplify the discussion, we neglect chemical dissociation of
the bubble content. The bubble radial dynamics is described by a
Keller equation \cite{kellerkolod,prosperlezzi1,brenner2002}. We study
the case of an air bubble in water, at ambient pressure
$p_0=101300$~Pa, driven at 20~kHz. We first consider a bubble in water
at ambient temperature $T_0 = 298$~K, with physical properties of
water $\rho_l=$~1000 kg/m$^3$, $\mu_l=$~10$^{-3}$ Pa.s, $\sigma
=$~0.0725 N.m$^{-1}$

\begin{figure}[h!tb]
  \includegraphics[width=\linewidth]{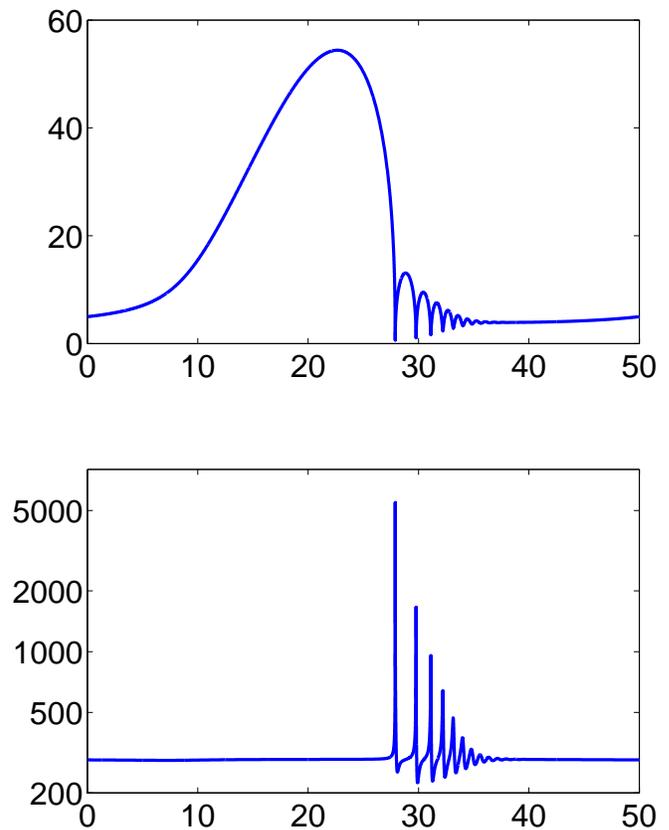}
  \caption{a. Temporal evolution of the radius of a 5 $\mu$m air
    bubble in water driven by a 20~kHz field at
    $p_a=130$~kPa. b. Temporal evolution of the bubble core temperature.}
    \label{figRT}%
\end{figure}

\begin{figure}[h!tb]
  \includegraphics[width=\linewidth]{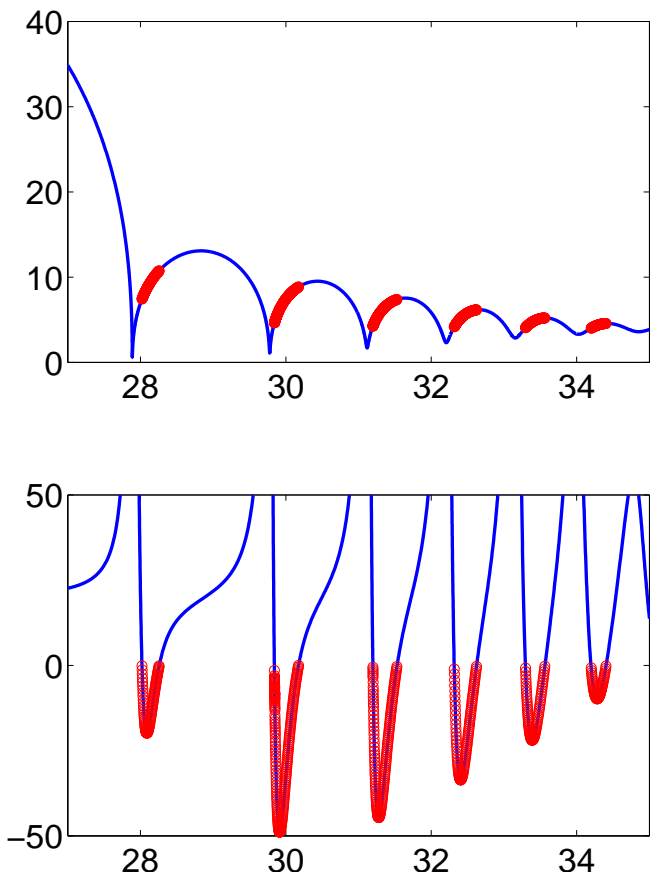}
  \caption{Zoom of Fig.~\ref{figRT} around the collapses. The red
    (color online) circle marks on both curves materialize the times
    at which the bubble core is at a temperature below 0$^\circ$C}
    \label{figRTzoom}%
\end{figure}

Figure~\ref{figRT} displays the variations of the radius
(Fig.~\ref{figRT}a) and core temperature (Fig.~\ref{figRT}b) of an air
bubble of ambient radius $R_0 = 5 \: \mu$m, driven by a 20~kHz
sinusoidal field of amplitude $p_a=130$~kPa. The classical temperature
peaks can be observed at the main and secondary collapses.  Moreover,
an interesting feature of the temperature curve can be observed on
Figs~\ref{figRTzoom}a-b, which are zoom into Figs~\ref{figRT}a-b
around the bubble afterbounces: slightly after the main collapse and
also after the secondary ones, the bubble core temperature falls down
\emph{below}~0$^\circ$C for a short time during the bubble
re-expansion.  Although this feature is visible in other works (see
for example Fig.~1c in~\cite{yasui96}, Fig.~1d in~\cite{yasui97},
Fig.~4 in \cite{gonghart98}, Fig.~2 in~\cite{kim2007}), to our
knowledge, it has never been commented.

The physical origin of this feature shares some similarities with the
adiabatic heating of the bubble core during the collapse: the
expansion velocity is fast enough to partially inhibit heat conduction
between the liquid and the bubble core, so that the expansion phases
are almost adiabatic.

\begin{figure}[h!tb]
  \includegraphics[width=\linewidth]{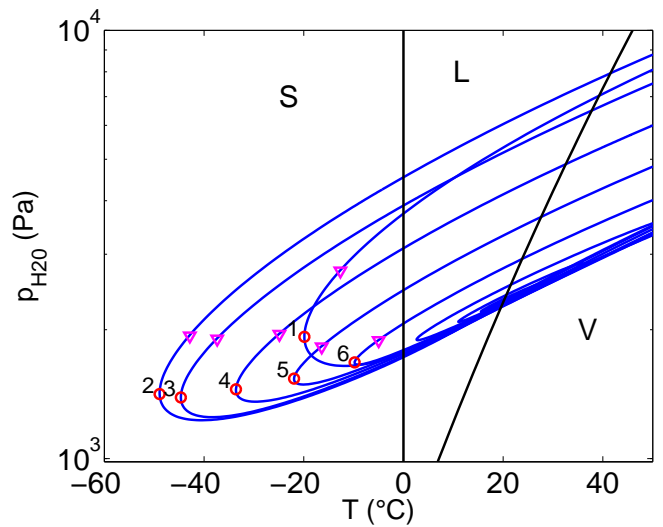}
  \caption{(Color online) Evolution of water conditions in the phase
    diagram of water $(T, \phdeuxo)$ in the conditions of
    Fig.~\ref{figRT} (blue thick solid lines). The phase boundaries
    are drawn in black thin dashed lines. The circles materialize the
    six negative peak temperatures visible in Fig.~\ref{figRTzoom},
    numbered by increasing time. The triangles materialize the water
    vapor state slightly before the minimal temperature is attained,
    and are represented to illustrate the direction followed by the
    trajectories. The dashed lines are the boundaries between the
    vapor, liquid and solid states of water.  The excursion of the
    trajectories in the solid region are, for the paths numbered from
    1 to 6: 236 ns, 332 ns, 331 ns, 305 ns, 266 ns, 204 ns.  }
    \label{figphase}%
\end{figure}

Moreover, owing to finite diffusion of vapor through air, there is
excess water trapped in the bubble core during the bounces, which
slowly condenses at the bubble wall from one bounce to the other
\cite{storeyszeri2000,toegel2000}.  Thus, when the temperature of the
bubble core drops below 0$^\circ$C, the water vapor content is cooled
down to freezing temperatures.  This raises the question of whether
this water vapor is thermodynamically stable against liquefaction or
ice formation. In order to assess this issue, we first plot the path
followed by the state of water in a temperature/water partial
pressure phase diagram.  Fig.~\ref{figphase} displays a subset of this
path around the post-collapse bubble expansions materialized by red
circle-marks (color online) on Fig.~\ref{figRTzoom}. It can be seen
that the vapor state crosses several time the liquid-vapor boundary
and even performs six excursions in the solid region, during time
periods of hundreds of ns.  This timescale, although very short
compared to the acoustic period, is much larger than the collapse
characteristic time.  The partial pressure of water vapor ranges
between $10^3$ and $10^4$ Pa, so that at this scale the liquid-solid
boundary is almost the vertical line $T=0^\circ$~C. The numbers
correspond to the order of appearance in the acoustic cycle, and the
paths are described anti-clockwise. The second excursion yields the
lowest temperature ($-48.9^\circ$~C), as could also be seen in
Fig.~\ref{figRTzoom}b.

%
%
The present simulation suggests therefore that even in a liquid at
ambient temperature, the water vapor of the bubble becomes repeatedly
metastable during the bubble bounces, not only against liquefaction,
but also against ice formation, for time lapses of about hundreds
of~ns.  This opens the interesting issue of whether this metastable
vapor has enough time to nucleate into droplets or ice crystals
\emph{inside the bubble}. Before going further, one should comment on
the singularity of the above results, in view of the parameters used
for the simulation yielding Figs.~\ref{figRT}-\ref{figphase}, which
are typical of bubbles levitated in SBSL cells. Such bubbles are known
to keep their spherical shapes for a large number of cycles, so that
they indeed undergo bounces. The present result would thus suggest
that inertial single bubbles commonly observed in levitation cells
have their content ready for ice nucleation in the expansive part of
their bounces, even at room temperature. Whether ice nuclei have
indeed enough time to form in such bubbles or not has therefore to be
clarified.

\begin{figure}[h!tb]
  \includegraphics[width=\linewidth]{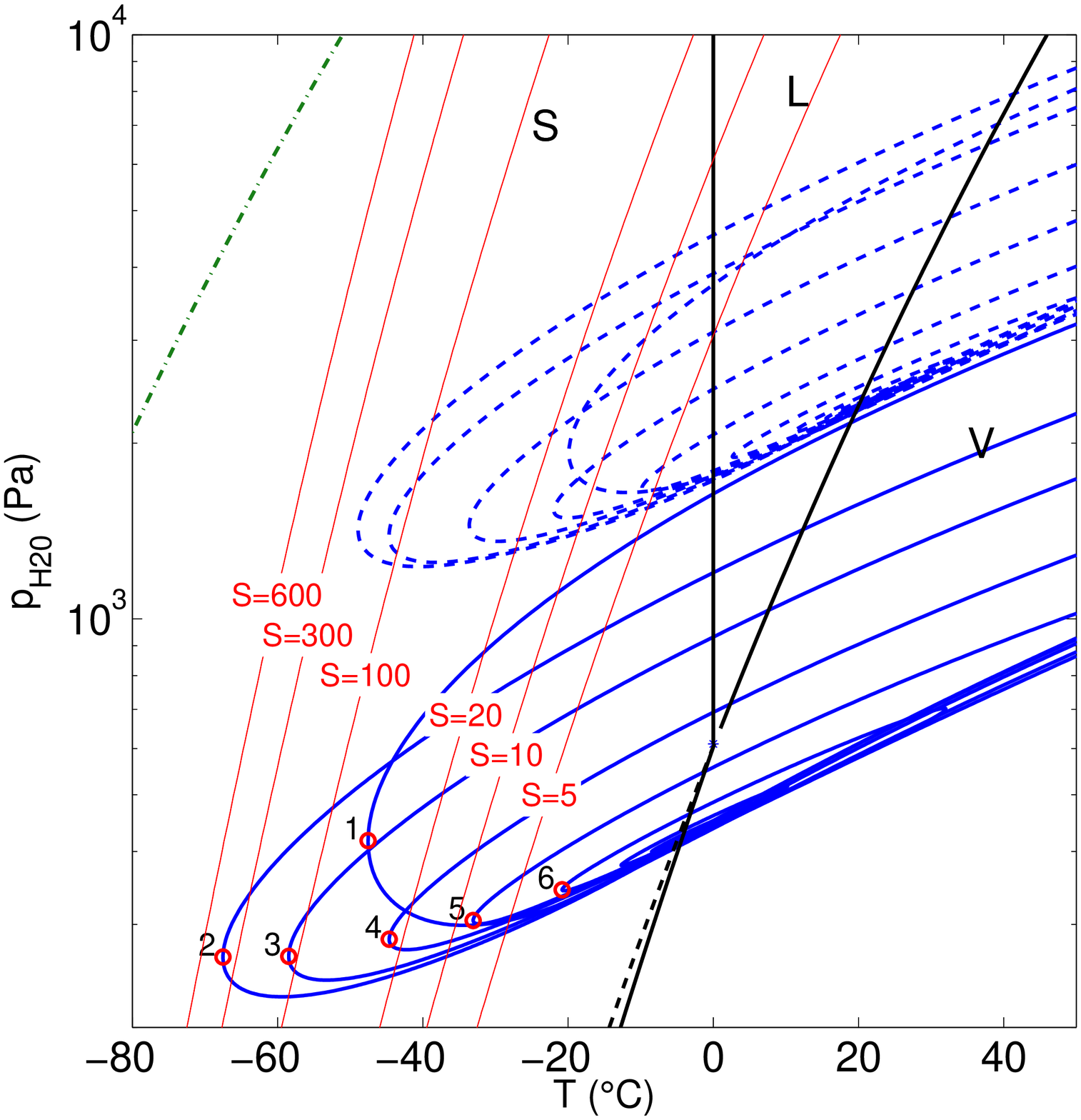}
  \caption{(Color online) Evolution of water conditions in the phase
    diagram of water $(T, \phdeuxo)$ for a~5~$\mu$m air bubble in
    water driven by a 20~kHz field at $p_a=130$~kPa. Solid (blue)
    line: $\tlneg$; Dashed (blue) line: $\tlamb$ (same as
    Fig.~\ref{figphase}). The phase boundaries are drawn in thin
    dashed (black) lines. The additional thin dashed line starting
    from the triple point is the equilibrium curve between supercooled
    water and vapor. There are 10 excursion loops in the solid whose
    respective durations are 908 ns, 663 ns, 531 ns, 455 ns, 420 ns,
    417 ns, 413 ns, 425 ns, 445 ns, and 514 ns. The (red) solid line
    are the supersaturation ratio level curves from
    Eq.~(\ref{defS}). The (green) dash-dotted line is an estimation of
    the vapor spinodal line (see appendix~\ref{annSpinodale})}
    \label{figphaseTambTneg}%
\end{figure}

If the latter observation also held for a bubble surrounded by
supercooled water, this result would provide a new explanation, albeit
incomplete, of how cavitation can trigger so easily ice formation in
supercooled water. In order to clarify this point, we reproduced the
precedent simulation in supercooled water at temperature $\tlneg$.
The path of the water state in the plane ($T$, $\phdeuxo$) is
displayed as a solid line in Fig.~\ref{figphaseTambTneg}. The path
obtained for $\tlamb$ (as displayed in Fig.~\ref{figphase}) is
recalled in dashed line for comparison.  Several differences can be
observed. As expected, the bubble inner temperature is shifted toward
low values since the surrounding liquid is colder. It is seen that the
minimal temperatures attained during bounces are thus lower than for a
bubble at ambient temperature (the minimum temperature attained
reaches $-67.5^\circ C$ in this case).  Moreover, the water vapor
pressure is also shifted toward lower values, since the bubble in colder
water initially contains less vapor.

Thus, in the case of a bubble surrounded by supercooled water, the
water vapor in the bubble is therefore not only metastable against
liquefaction, but moreover enters in the temperature range below the
so-called \textit{homogeneous freezing point of water} at -42$^\circ
C$, which has long been considered as the supercooling limit for
liquid water \cite{Angell83}, below which homogeneous freezing occurs.
One might therefore be tempted to conclude that ice should nucleate,
at least temporarily in the bubble during the excursions in the solid
phase on Fig.~\ref{figphaseTambTneg}. Whether freezing indeed occurs
inside the bubble is the main question discussed in this paper. We
first propose a short review of the abundant literature on water
freezing, notably in the context of cloud microphysics, and extending
in the more general frame of water physics.

%
\section{Review of water freezing }
%

\subsection{Cloud physics}
The conditions in the bubble are similar to the ones encountered in
clouds, and the related issue of the formation of liquid and/or ice in
the latter. The state of water in clouds has been recently found to be
crucial for climate change predictions, and motivated an active
research in this field. Clouds are formed by expansion and cooling of
humid air as it ascends, so that vapor becomes thermodynamically
unstable and condensation into liquid droplets occurs
\cite{Debenedetti1996book,Mason2010Book}. This condensation is
favored by the presence of soluble or insoluble aerosol particles
called cloud condensation nuclei (CCN) \cite{Mason2010Book}. Formation
of ice can occur at moderate negative temperatures by several
mechanisms involving specific aerosol particles (called ice forming
nuclei, IFN) in low and middle troposphere clouds
\cite{CantrellHeymsfield2005,MurrayOSullivan2012}.  However in high
altitude clouds such as cirrus, the dearth of such IFNs prevents such
mechanisms and supercooled water droplets can be found down to
$-40$~$^\circ$C
\cite{SassenLiou85,Sassen92,Heymsfield93,Debenedetti2003}. Below this
temperature, the supercooled droplets in the cloud freeze
homogeneously.

\subsection{Water no man's land}
These results are in close agreement with laboratory experiments,
which have evidenced the so-called \textit{homogeneous freezing
  temperature} $T_H$ below which supercooled droplets unavoidably
undergo homogeneous freezing. According to classical nucleation
theory, this temperature in fact slightly varies with the droplet
size, since the number of nucleated embryos per unit time is
proportional to the sample volume. At the time of Angell's early
review on supercooled water \cite{Angell83}, the lowest temperature
ever reached for supercooled pure water was -42$^\circ$C in
micron-sized droplets \cite{mossop55}.  

However, it has been demonstrated that ultra-fast cooling of water
allows to bypass homogeneous freezing, and to obtain amorphous ice with
a glass transition at 136~K, which further crystallize into cubic ice
(Ic) near $150$~K upon heating \cite{Debenedetti2003}. There exists
therefore a domain between 150~K and 231~K, termed as ``\textit{no
  man's land}'' \cite{MishimaStanley98}, where the existence and
properties of supercooled water could not be assessed for long. 

Entering the no man's land has been achieved first by the pioneering
work of Bartell and co-workers \cite{BartellHuang94,HuangBartell95},
by expanding a mixture of carrier gas (neon) and water vapor through
small supersonic nozzles. With cooling rates of $10^7$~K/s, liquid
water clusters of 74~\AA{} (6000 molecules) could be observed and
analyzed by electron diffraction spectroscopy. The clusters were found
to freeze at temperatures as low as 200~K, and diffraction patterns
showed that cubic ice (Ic) was nucleated. Similar recent experiments
with slightly lower cooling rate ($10^5$~K/s) and argon as the carrier
gas allowed to nucleate and freeze supercooled droplets between 202~K
and 215~K \cite{Manka2012}. The estimated homogeneous nucleation rates
in supersonic nozzle experiment can reach about
$10^{30}$~m$^{-3}$s$^{-1}$, which is between 15 and 20 orders of
magnitude larger than the nucleation rates observed slightly above the
no man's land, where numerous results have been collected
\cite{HuangBartell95}. In spite if this large range of nucleation
rate, models based on classical nucleation theory, and assuming
nucleation of cubic ice (Ic), seem to yield rather good results, even
if there remains uncertainties on the interfacial energy between
liquid and ice (Ic) \cite{Murray2010,Manka2012}.

That deeply supercooled water freezes into cubic ice (Ic) rather than
hexagonal ice (Ih) is supported by Bartell's electron diffraction
spectra \cite{BartellHuang94,HuangBartell95} and by the presence of
cubic ice in clouds \cite{Riikonen2000halo,Goodman89}. As noted by
Murray and co-workers \cite{MurrayOSullivan2012}, this also agrees
with Ostwald's rule of stages, which states that the metastable phase
nucleate (in this case cubic ice) preferentially to the stable phase
(hexagonal ice). Recent experiments, computer simulations, and careful
reinterpretations of past studies, revealed that ice formed from
supercooled water is in fact composed of randomly stacked layer of
cubic and hexagonal sequences \cite{MalkinMurray2012}. As noted by
Murray and co-workers \cite{MurrayOSullivan2012}, the precise phase of
the ice critical nucleus remains unknown, and this constitutes an
additional difficulty in the establishment of a definitive nucleation
theory for supercooled droplets freezing, especially for the
estimation of the liquid-solid interfacial energy, which has a huge
influence on the nucleation rate. Despite the latter reservations, the
freezing rate proposed in Ref.~\cite{Murray2010} seems to yield good
agreement with a large set of experimental results, over a large range
of supercooling, including the no man's land \cite{Manka2012}.

As a final remark on ice nucleation, there seems to have general
evidence that direct homogeneous deposition of ice from vapor does not
occur, neither in clouds, nor in supersonic expansions experiments
\cite{Manka2012}. Thus, liquid droplets would always nucleate prior to
freezing, which, again, is a consequence of Ostwald's rule of stages.
It sounds therefore reasonable to discard such a mechanism in the
present case, and assume that the water vapor in the bubble cannot
form ice nuclei without the prior formation of liquid droplets.

%
\subsection{Relevance to the present problem}
%
Following one of the paths visible on Fig.~\ref{figphaseTambTneg}, it
is seen that the water vapor in the bubble core can be super-cooled
to temperatures falling in the no man's land range. If direct ice
deposition from vapor can be discarded, ice formation in the bubble
core, if any, could only occur by primary condensation of vapor into
droplets, which, further supercooled as the bubble expands, might
undergo homogeneous freezing, possibly in the no man's land region. A
reasonable theory for the latter process is available over a wide
range of freezing temperatures \cite{Murray2010,Manka2012}. The
physics involved is therefore strikingly similar to the one
encountered in supersonic nozzles phase transitions. In both cases,
the fast cooling of the mixture of water vapor and carrier gas results
from an almost adiabatic expansion. The latter owes to a supersonic
flow in the nozzle divergent in one case, and by the bubble outward
motion in the other.


There remains however the problem of the nucleation of droplets
itself. Ultrasonic nozzle experiments \cite{Manka2012} show that
droplets first nucleate when supersaturation $\phdeuxo/\psatliq(T)$
reaches a critical value and further grow by condensation of the
surrounding vapor. In doing so, heat is released in the carrier gas
and increases its temperature, which in turn quenches liquid
nucleation, yielding an almost monodisperse aerosol. Once water vapor
has almost entirely been consumed, the subsequent expansion in the
nozzle cools the droplets until they freeze homogeneously. During the
growth phase, droplets are hotter than the surrounding mixture, as
vapor condenses at their surface.

Models for this three-step process, droplet nucleation / droplet
growth / droplet freezing, are now available and show good agreement
with nozzle experiments
\cite{Hill1966Condensation,Bartell90,SinhaWyslouzil2009,Tanimura2010}.
Transposing such models to the present case is technically feasible,
by supplementing the ODE set describing the bubble motion with the
ODEs describing the droplets nucleation and growth. This would require
however to reconsider the thermal model for the bubble interior and
add a source term accounting for the heat released by vapor
condensation. Embarking in such a procedure is out of the scope of the
present paper.

Nevertheless, it is interesting to investigate at least the first step
along a given bubble bounce, to assess whether a significant number of
droplets can be nucleated, their sizes, and the fraction of the vapor
available in the bubble core that is condensed into clusters.  Before
carrying out these nucleation calculations, we must make sure that
spinodal liquefaction of water vapor in the bubble can be discarded in
the present case. As can be seen on Fig.~\ref{figphaseTambTneg}, the
paths followed by water in the $(T, \phdeuxo)$ plane remain
sufficiently far from the estimated vapor spinodal line (dash-dotted
line), whose estimation is deferred to
appendix~\ref{annSpinodale}. This ensures that condensation of liquid
water in the bubble, if any, can only occur through homogeneous
droplet nucleation.

%
\section{Kinetics of droplets nucleation}
%

%
\subsection{Nucleation model}
%
The instantaneous supersaturation ratio related to the vapor-liquid
transition is classically defined by:
\begin{equation}
  \label{defS}
  S(\phdeuxo, T) = \frac{\phdeuxo}{\psatliq(T)}
\end{equation}
where $\psatliq(T)$ is the equilibrium vapor pressure at temperature
$T$. As we are dealing with potentially supercooled water, the
correlation used for the latter must extend into the deeply
supercooled regime, and we use the results of
Ref.~\cite{MurphyKoop2005}. On the other hand, it should be emphasized
that $\phdeuxo$ is not constant in our case, because the bubble
continuously exchanges water with the surrounding liquid phase through
its interface. This is why, contrarily to most studies, the
supersaturation ratio does not depend on $T$ only.

\begin{figure}[h!tb]
  \includegraphics[width=\linewidth]{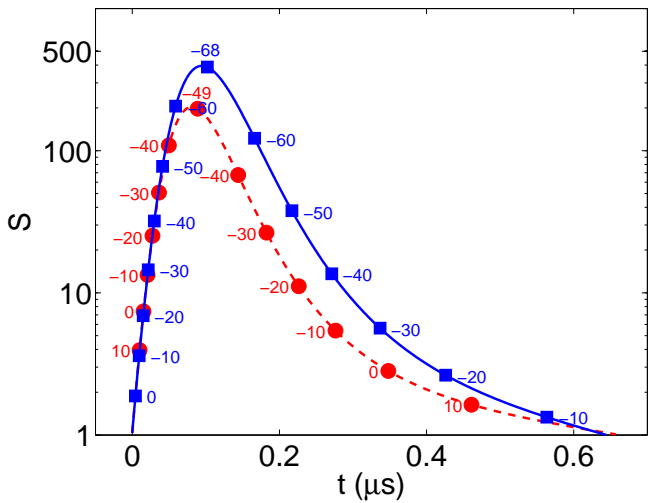}
  \caption{(Color online) Supersaturation ratio in a~5~$\mu$m air
    bubble in water driven by a 20~kHz field at $p_a=130$~kPa, along
    the second bounce. The conditions are the ones of
    Fig.~\ref{figphaseTambTneg}.  The origin of time is chosen when
    the water vapor state crosses the liquid-vapor equilibrium curve
    in each case. The markers on the curves indicate the bubble core
    temperature in $^\circ$C.}
    \label{figS}%
\end{figure}

The evolution of $S$ on the second rebound is displayed on
Fig.~\ref{figS} in the two cases corresponding to
Fig.~\ref{figphaseTambTneg}: $\tlneg$ ( solid line, blue online) and
$\tlamb$ (dashed line, red online). We also displayed the iso-S curves
in the ($T, \phdeuxo)$ plane in Fig.~\ref{figphaseTambTneg} (red online).

The classical nucleation theory (CNT) \cite{kashiev} assumes that
nucleation occurs through progressive build-up of clusters of
molecules, that are precursors of the new phase. A $\gclust$-sized
cluster can yield a $(\gclust+1)$-sized cluster by attachment to a
water molecule (termed as ``monomer''), and conversely the latter can
loose a monomer by the backward reaction:
\begin{equation}
  \label{nuclreac}
  (\gclust) + (1) 
  \renewcommand{\arraystretch}{0.1}
  \begin{array}{c}
    \rightharpoonup \\ \leftharpoondown
  \end{array}
 (\gclust+1), \qquad g = 1, 2, \dots
  \renewcommand{\arraystretch}{1}
\end{equation}

Under the so-called capillary approximation, the formation energy of a
$\gclust$-sized cluster can be written as
\begin{equation}
  \label{defWtheta}
  W(\gclust) = k T \left(\Theta \gclust^{2/3} - \gclust\ln S \right), \qquad   
  \Theta = \frac{a\sigma}{k T}
\end{equation}
where $a$ is the area of a monomer, $\sigma$ is the interfacial free
energy between the bulk phases, and $\Theta$ its dimensionless form.
Assuming spherical clusters, $a=(36\pi v_0^2)^{1/3}$, where $v_0$ is
the molecular volume.

The energy of formation has a maximum $\wet$, which is the nucleation
barrier, for a cluster of critical size $\net$, defined by :
\begin{equation}
  \wet = \frac{4}{27}  \frac {\Theta^3} {(\ln S)^2} k T\,
  \qquad
  \net = \left( \frac{2}{3} \frac{\Theta}{\ln S} \right)^3,
\end{equation}
so that the nucleation rate is essentially the rate of production of
critical clusters. Assuming that the cluster size $g$ is a continuous
variable, the kinetics of the reaction set~(\ref{nuclreac}) can be
described by a standard balance equation:
\begin{equation}
  \label{mastereq}
  \dsurd{\csd(\gclust,t)}{t} = -\dsurd {J(\gclust, t)} {\gclust},
\end{equation}
where $\csd(\gclust, t)$ is the concentration in $\gclust$-sized
clusters and $J(\gclust, t)$ is the cluster flux along the cluster
size axis~$(\gclust)$. This flux can be shown to read:
\begin{equation}
  \label{defJclust}
  J(\gclust, t) =  - \ratt(\gclust, t) \csd(\gclust, t) 
  \dsurd {} {g} \left[
    \frac{ \csd(\gclust, t) }{ \csdeq(\gclust, t) }
     \right],
\end{equation}
where $\ratt(\gclust, t)$ is the attachment rate by collisions between
monomers in concentrations $\csd(1,t)$ and a $\gclust$-sized cluster:
\begin{equation*}
  \ratt(\gclust,t) = \csd(1,t) a g^{2/3} 
  \left(\frac{k T}{2\pi m} \right)^{1/2}
\end{equation*}
with $m$ mass of a water molecule. The quantity $\csdeq(\gclust, t)$
is the so-called equilibrium concentration of clusters deduced from
the law of mass action.

The correct expression for the latter has been a matter of debate
\cite{kashiev} since the initial work of Becker and D\"oring
\cite{BeckerDoring35}.  We use the self-consistent expression proposed
by Girshick and co-workers
\cite{Girshick90kinetic,Girshick90TimeDependent,Kashchiev2006Analysis},
which, contrarily to the classical formulation, has the advantage to
be valid for~$\gclust=1$:
\begin{equation}
  \label{defclusteq}
  \csdeq(\gclust,t) = \csdeq(1,t) 
  \exp \left[ - \frac {W(\gclust)-W(1)} {k T} \right]
\end{equation}

The set of equations (\ref{mastereq})-(\ref{defJclust}) is known as
master equation of nucleation. It has no analytic solution in the
general case of time-dependent supersaturation, as in the present
problem. In particular, as far as we are aware, analytic treatment
of the number of nuclei produced by a supersaturation pulse has been
poorly explored. A noticeable exception can be found in the work of
Trinkaus~\&~Yoo \cite{TrinkausYoo87} who used Green functions
formalism to derive approximate analytic solutions of the master
equation, in the case of an idealized nucleation barrier whose
location has a parabolic time-dependence around a minimum $\net(t)$.
However, the analytic expressions proposed by these authors is
restricted to the case where the system is still supersaturated when
the supersaturation pulse has relaxed, which is not the case here. We
must therefore revert to some approximation. The  two solutions
adopted are described hereafter.

We first note that the simpler problem of transient nucleation in
response to a supersaturation step has been extensively studied
\cite{Kashchiev70_Transient,ShiSeinfeld90Transient,%
  DemoKozisek93_homogeneous,kashiev}. In this case, the transient
duration is of the order of the so-called nucleation time lag:
\begin{equation}
  \label{timelag}
  \tau = \frac{\delta^2}{2 \ratt(\net)},
\end{equation}
where 
\begin{equation}
  \label{defdelta}
  \delta = \left. \left(
      -\frac{1}{2 k T} \ddesurd{W}{\gclust}
    \right)^{-1/2}\right|_{\gclust = \net}
  = 3 {\net}^{2/3} \Theta^{-1/2}
\end{equation}
is the width of the nucleation barrier. The time-lag is physically the
order of magnitude of the time required for the clusters to populate
the subcritical region $\gclust < \net$ by fluctuations.  

%
\subsection{Quasi stationary nucleation rate}
%
If supersaturation has been maintained constant for a duration
sufficiently larger than the time-lag, nucleation can be considered
stationary, and $\sdsurd{\csd(\gclust,t)}{t} = 0$.  Under some
approximations, Eq.~(\ref{mastereq}) can then be solved to obtain the
steady-state cluster concentration \cite{kashiev}. The stationary
nucleation rate $J_S$ is defined as the cluster flux through the
critical size $J(\net)$ and is found to write:
%
%
\begin{eqnarray}
  \nonumber
  J_S &=& \left(\frac{2\sigma}{\pi m} \right)^{1/2}
  \left(\frac{\psatliq(T)}{k T} \right)^2   v_0 e^\Theta S  \\
  \label{defJstat}
  & &\quad \times
  \exp\left( -\frac{4}{27} \frac{\Theta^3}{(\ln S)^2} \right).
\end{eqnarray}
%

\begin{figure}[h!tb]
  \includegraphics[width=\linewidth]{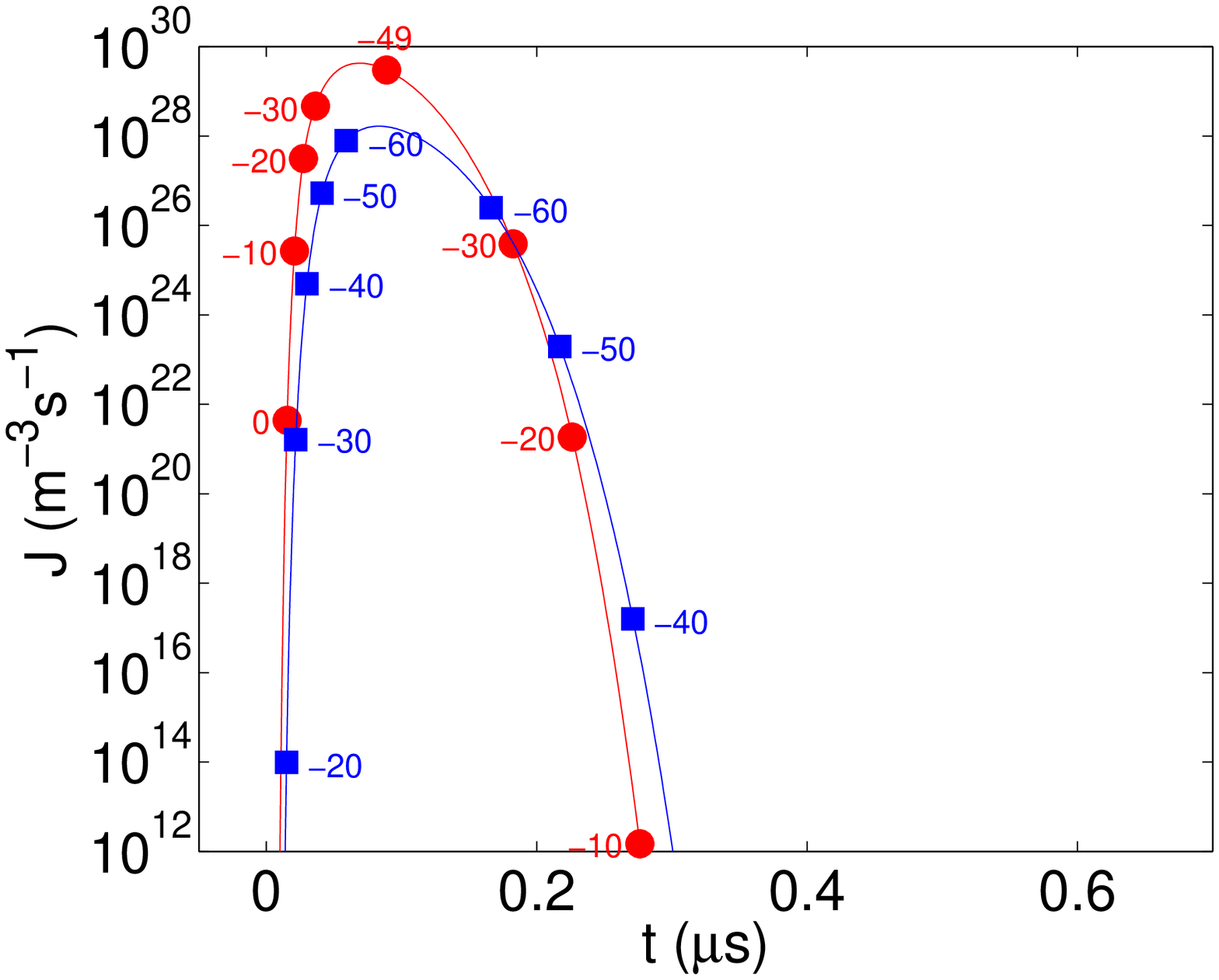}
  \caption{(Color online) Stationary nucleation rate of droplets in
    a~5~$\mu$m air bubble in water driven by a 20~kHz field at
    $p_a=130$~kPa, along the second bounce, calculated from
    Eq.~(\ref{defJstat}).  Square symbols (blue online): $\tlneg$;
    Round symbols (red online): $\tlamb$ (same as
    Fig.~\ref{figphase}). The origin of time is chosen when the water
    vapor state crosses the liquid-vapor equilibrium curve in each
    case. The markers on the curves indicate the bubble core
    temperature in $^\circ$C.}
    \label{figJ}%
\end{figure}

We calculated the latter quantity along the second bounce, in the
conditions of Fig.~\ref{figphaseTambTneg}, for a liquid at ambient
temperature and for a supercooled liquid at $\tlneg$
(Fig.~\ref{figJ}). In both cases, the nucleation rate increases
sharply and reaches a maximum after about 100~ns after entering in the
metastable liquid zone. The main difference is that both the
nucleation rate and the temperature drop much more slowly in the
bubble surrounded by the supercooled liquid.  On the other hand, a
counter-intuitive result is that the maximum nucleation rate is lower
for the bubble in the supercooled liquid.  This is mainly due to the
pre-factor $(\psat/(k T))^2$ which is lower for the supercooled bubble
because there is less vapor in the latter (see
Fig.~\ref{figphaseTambTneg}).

As proposed by Kashchiev
\cite{Kashchiev70_S_de_t,kashiev,TrinkausYoo87}, the nucleation rate
Eq.~(\ref{defJstat}) can still be used if supersaturation evolves on a
time scale much larger than the time-lag, a situation referred to as
quasi-stationary nucleation. Assuming this assumption valid in our
case, the number of critical nuclei formed at time $t$ can be
expressed as:
\begin{equation}
  \label{defNquasi}
  \Nclustquasi(t) = \int_0^t J_S \bigl[S(s), T(s)\bigr]\, V(s) \; d s, 
\end{equation}
whereas the number of water molecules condensed into nuclei reads:
\begin{equation}
  \label{defnquasi}
  \nquasi(t) = \int_0^t J_S \bigl[ S(s), T(s)\bigr]\, V(s)\, \net(s) \; d s, 
\end{equation}
In both integrals, the origin of time is chosen when the vapor
becomes supersaturated.

In the present case, calculated instantaneous values of the time lag
$\tau$ are found to be larger than 1~$\mu$s, except near the
supersaturation peak where it becomes of the order of 100 ns. It has
therefore the same order of magnitude as the typical time scale of the
supersaturation variations, so that the assumption of quasi-stationary
nucleation may be not fully reliable here.  For a given change of the
supersaturation value, nucleation does not reach steady state, and the
use of Eqs.~(\ref{defNquasi}) probably yields an overestimation of the
number of nuclei formed.

%
\subsection{Transient nucleation}
\begin{figure}[h!tb]
  \includegraphics[width=\linewidth]{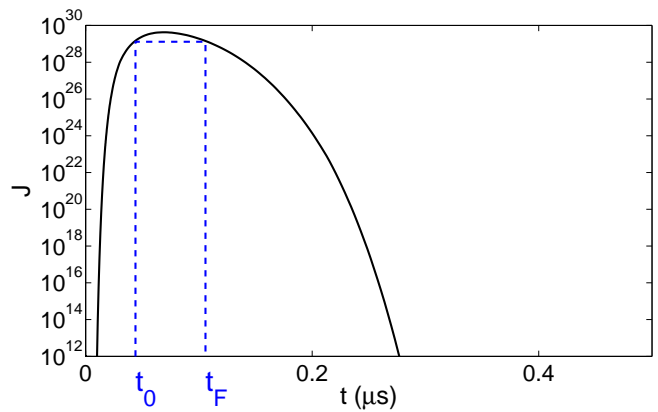}
  \caption{Choice of start-time $t_0$ and end-time
    $t_F$ to apply the transient nucleation results of
    \cite{ShiSeinfeld90Transient}. The values are chosen so that the
    stationnary nucleation rate $J_S$ lies in the range
    $\alpha J_S^\text{max} \leq J_S \leq J_S^\text{max}$, where
    $\alpha$ is a free parameter of order 1 (typically $\alpha=1/2$,
    see text and Tab.~\ref{tabresults}) }
    \label{figpulse}%
\end{figure}

%
The second method used is to revert to some treatment of transient
nucleation in response to a supersaturation step.  In the present
case, supersaturation evolves rather as a dome-shaped pulse, and a
turnaround must be used. Since the rise of $S(t)$ is very abrupt, we
replace it by a square pulse between $\tinit$ and $\tend$, defined as
(see Fig.~\ref{figpulse}).  The latter times are chosen such that
$J_S(\tinit) = J_S(\tend) = \alpha J_S^\text{max}$, where $\alpha$ is
a free parameter.

If we restrict our analysis to times lower than $\tend$, all happens
as if the system had undergone a supersaturation step at $\tinit$, and
classical results of transient nucleation can be used up to time
$\tend$. The main drawback of the method is that the choice of times
$\tinit$ and $\tend$ is somewhat arbitrary, and the sensitivity of the
results to this parameter will be examined a posteriori.


We then used the results of Shi, Seinfeld \& Okuyama
\cite{ShiSeinfeld90Transient}, who used a combination of a
boundary-layer method and Laplace transform to solve the nucleation
master equation (\ref{mastereq}) for a supersaturation step at $t=0$.
They obtained an analytic expression of the instantaneous cluster
size distribution \footnote{We note that there is an error in Eq.~(25)
  in the original paper of Shi, Seinfeld \& Okuyama
  \cite{ShiSeinfeld90Transient}, which yields inconsistent results at
  $t=0$. This error has been commented and corrected in
  Ref.~\cite{ShiSeinfeld94_Universal}.}:
\begin{eqnarray}
  \nonumber
  \frac{\csd(\gclust, t)} {\csdeq(\gclust)} &=& \frac{1}{2}
  \textrm{erfc} \left[ 
    \frac{\gclust - \net}{\delta} + \exp\left(-\frac{t}{\tau} + \lambda \right)
  \right] \\
  & & \qquad - \frac{1}{2}
  \textrm{erfc} \left[ 
    \frac{\gclust - \net}{\delta} + e^\lambda 
  \right],
  \label{yshi}
\end{eqnarray}
where 
\begin{equation}
  \label{deflambda}
  \lambda = {\net}^{-1/3} - 1 +
  \ln\left[3\net\frac{1-{\net}^{-1/3}}{\delta}
  \right].
\end{equation}
The concentration of nuclei produced
\begin{equation*}
  \cnucl(t) = \int_0^t J(\net, s) \; d s
\end{equation*}
was also obtained in analytic form by the authors as:
\begin{equation}
  \label{defcnuclshi}
  \cnucl(t) = J_S \frac{\tau}{2} \left[
    E_1 ( e^{ 2 (\lambda-t/\tau) }) - E_1(e^{2\lambda}) \right],
\end{equation}
where $J_S$ is the stationary nucleation rate given by
Eq.~(\ref{defJstat}), $\tau$ is the nucleation time lag from
Eq.~(\ref{timelag}), and $E_1$ is the exponential integral.

In order to use these analytic results, we set the quantities
$\net$, $\delta$ and $\tau$ to their values at $\tinit$
and replace $t$ by $t-\tinit$, up to $\tend-\tinit$.  The total number
of nuclei $\Nclust(t) = \int_{\tinit}^{\tend} J(s) V(s) d s$ formed in
the bubble was then approximated by:
\begin{equation}
  \label{defNtransient}
  \Nclusttransient (t) \simeq \cnucl(t-\tinit)  V_m, 
  \qquad \tinit < t < \tend
\end{equation}
with $\cnucl$ given by Eq.~(\ref{defcnuclshi}), and $V_m$ is the mean
volume of the bubble during the equivalent supersaturation pulse
between $\tinit$ and $\tend$. The number of water molecules condensed
into critical nuclei reads similarly:
\begin{equation}
  \label{defntransient}
  \ntransient (t) \simeq \cnucl(t-\tinit)  V_m \net, 
  \qquad \tinit < t < \tend
\end{equation}
Since we replaced the exact dome-shaped supersaturation pulse by a
smaller square pulse, these estimations are expected to yield lower
bounds of the number of nuclei produced and the number of molecules
consumed, respectively.

%
\subsection{Results}
%

\begin{figure}[h!tb]
  \includegraphics[width=\linewidth]{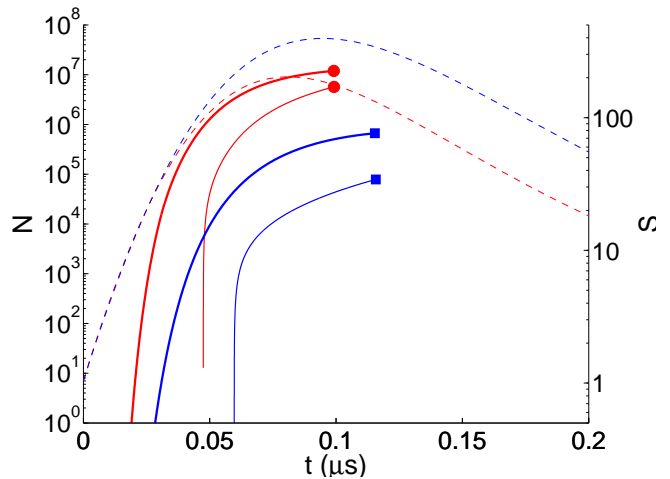}
  \caption{(Color online) Number of nuclei produced during the
    supersaturation pulse; Thick solid lines: assuming
    quasi-stationary nucleation ; Thin solid lines: assuming transient
    nucleation.  The conditions are the same as in fig.~\ref{figJ}.
    Square symbols (blue online): $\tlneg$; Round symbols (red
    online): $\tlamb$ (same as Fig.~\ref{figphase}). The
    supersaturation is recalled in dashed lines (right ordinate axis)
    for the two conditions.}
    \label{figN}%
\end{figure}

Figure \ref{figN} displays the number of nuclei produced,
$\Nclusttransient(t)$ from Eq.~(\ref{defNtransient}) (thin solid
lines) and $\Nclustquasi(t)$ from (\ref{defNquasi}) (thick solid
lines), in the conditions of Figs.~\ref{figphaseTambTneg}-\ref{figJ}. The
curves for the bubble in ambient water $\tlamb$ end with round symbols
(red line online), and the one in supercooled water $\tlneg$ with
square symbols (blue online). The times $\tinit$ and $\tend$ have been
chosen on the criterion $J_S(\tinit) = J_S(\tend) = \frac{1}{2}
J_S^\text{max}$. The two estimations are expected to yield lower and
upper boundaries, respectively, of the number of liquid nuclei formed
in the bubble. The number of nuclei obtained at $\tend$ in the two
cases are displayed as bold lines in Tab.~\ref{tabresults}.  As
expected from the comparisons of the nucleation rates on
Fig.~\ref{figJ}, less nuclei are produced in the bubble surrounded by
supercooled water.

We can also make use of Eqs.(\ref{defnquasi})-(\ref{defntransient}) to
calculate upper and lower bounds for the number of water molecules
that have clusterized into nuclei, or more eloquent, the fraction~$x$
of those molecules relative to the initial number of water molecules
in the bubble as the vapor becomes supersaturated (rightmost columns
of Tab.~\ref{tabresults}). It can be seen that during the second
rebound of the bubble surrounded by liquid at ambient temperature,
between 5~\% and 10~\% of the water vapor initially present can
clusterize into droplets. This fraction drops down to between 0.35~\%
and 2.5~\% in the case of a bubble in supercooled water.

\begin{table}[htp]
  \centering
  \begin{tabular}{cccccccc}
    \hline
    $R_0$ & $p_a$ & $\tlinfty$ & 
    $\displaystyle\alpha = \frac{J_\text{cut}}{J_\text{max}}$ & 
    $\Nclusttransient$ & $\Nclustquasi$ & 
    $\xtransient$ & $\xquasi$ \\
    $\mu$m & kPa & $^\circ$C &  & 
    & & \% & \% \\
    \hline
      \textbf{5} & \textbf{130} & \textbf{-5} & \textbf{1/2} & 
  $\mathbf{\puisdix{  7.8}{4}}$& $\mathbf{\puisdix{  6.7}{5}}$ &     
  \textbf{0.35} & \textbf{2.5} \\ 
  5 & 140 & -5 & 1/2 & $\puisdix{  7.5}{4}$& $\puisdix{  6.5}{5}$ &      0.32 &       2.5 \\ 
  5 & 150 & -5 & 1/2 & $\puisdix{  7.1}{4}$& $\puisdix{  6.4}{5}$ &      0.31 &       2.4 \\ 
  \textbf{5} & \textbf{130} & \textbf{20} & \textbf{1/2} & 
  $\mathbf{\puisdix{  5.7}{6}}$& $\mathbf{\puisdix{  1.2}{7}}$ & 
  \textbf{5.6} &       \textbf{10.} \\ 
  5 & 140 & 20 & 1/2 & $\puisdix{  4.9}{6}$& $\puisdix{  1.1}{7}$ &       4.7 &       9.3 \\ 
  5 & 150 & 20 & 1/2 & $\puisdix{  4.2}{6}$& $\puisdix{  1.0}{7}$ &       3.9 &       8.3 \\ 

      5 & 130 & -5 & 1/3 & $\puisdix{  6.5}{4}$& $\puisdix{  7.2}{5}$ &      0.32 &       2.8 \\ 
  5 & 140 & -5 & 1/3 & $\puisdix{  6.5}{4}$& $\puisdix{  7.1}{5}$ &      0.31 &       2.7 \\ 
  5 & 150 & -5 & 1/3 & $\puisdix{  5.9}{4}$& $\puisdix{  6.9}{5}$ &      0.28 &       2.6 \\ 
  5 & 130 & 20 & 1/3 & $\puisdix{  5.0}{6}$& $\puisdix{  1.3}{7}$ &       5.5 &       11. \\ 
  5 & 140 & 20 & 1/3 & $\puisdix{  4.5}{6}$& $\puisdix{  1.2}{7}$ &       4.7 &       9.9 \\ 
  5 & 150 & 20 & 1/3 & $\puisdix{  4.2}{6}$& $\puisdix{  1.1}{7}$ &       4.2 &       9.0 \\ 

    \hline 
  \end{tabular}
  \caption{
    Number of clusters created and number of water molecules consumed
    during the buble second rebound, for various bubble ambient radii 
    and driving pressure amplitudes. The calculation have been 
    performed twice for each parameter set, using $\alpha=1/2$ and 
    $\alpha=1/3$, respectively (see Fig.~\ref{figpulse}). The two bold
    lines correspond to the data displayed on Fig.~\ref{figN}
  }
  \label{tabresults}
\end{table}

Table~\ref{tabresults} also displays results obtained for larger
driving acoustic pressures. It can be seen that increasing the driving
decreases the number of nuclei formed. This is due to the fact that
larger driving amplitudes produce hotter collapses, so that the
initial temperature at the beginning of the rebounds becomes larger.
Thus, vapor supersaturation decreases and so does the nucleation
rate.
 
Finally, we also varied the arbitrary factor $\alpha$ used to define
the equivalent supersaturation step (see Fig.~\ref{figpulse}) in order
to assess the sensitivity of the results to this quantity.
Table~\ref{tabresults} shows that the latter is reasonably weak.

\begin{figure}[h!tb]
  \includegraphics[width=\linewidth]{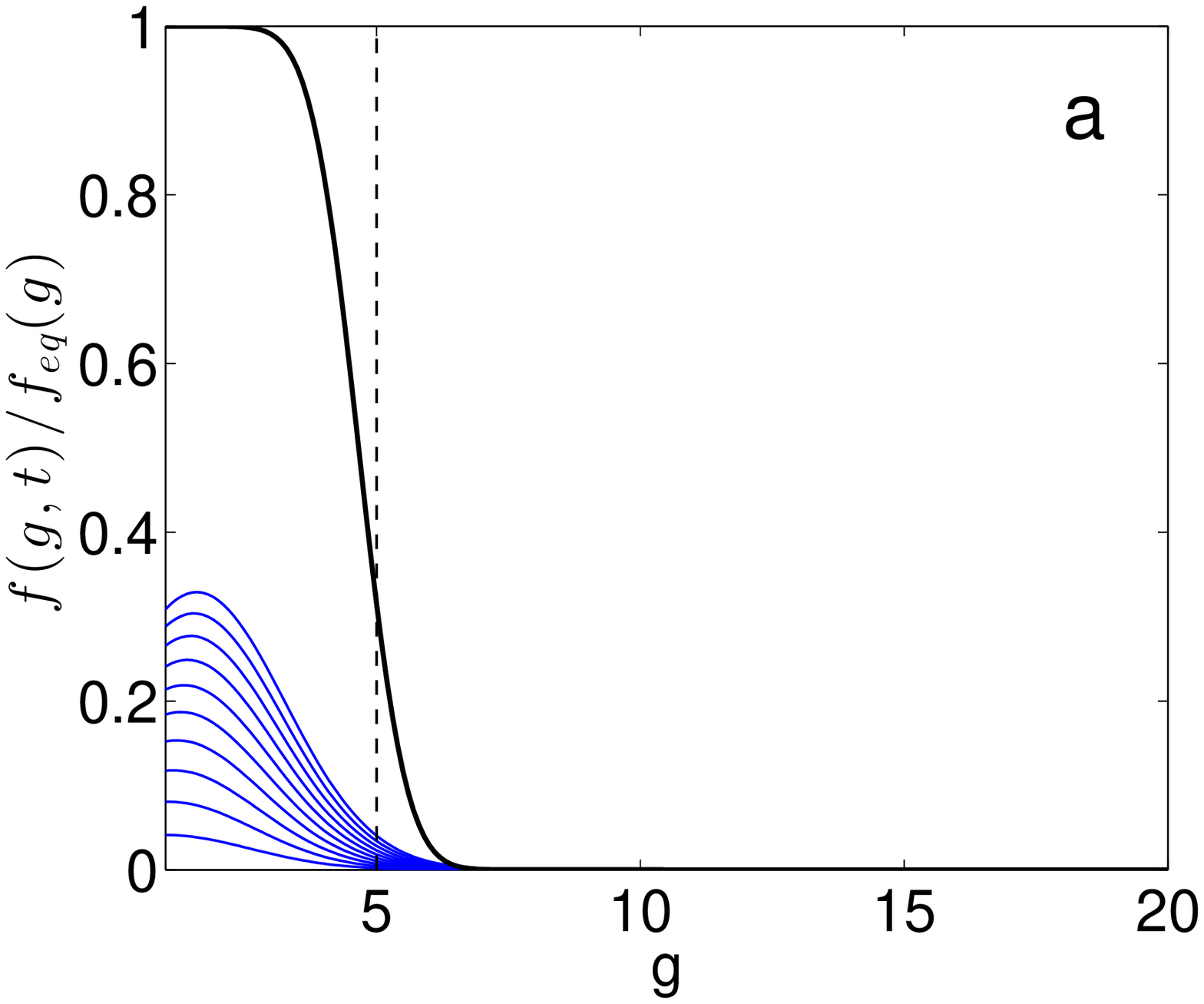}
  \includegraphics[width=\linewidth]{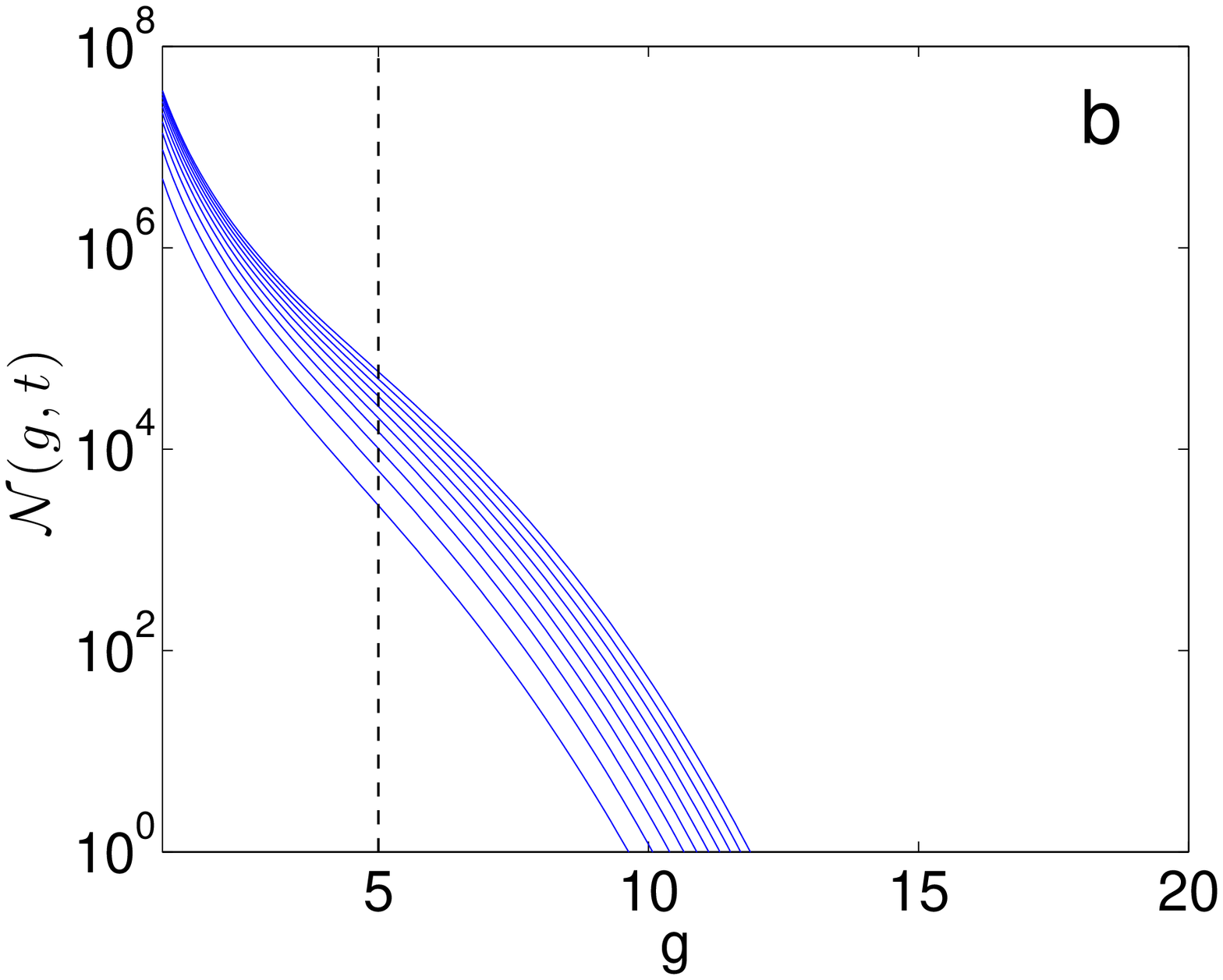}
  \caption{a. Thin lines: dimensionless cluster size distribution
    calculated from~(\ref{yshi}) at ten equidistant times covering the
    supersaturation pulse (time increasing from bottom to top). The
    thick line represents the number of clusters formed in steady
    state.  b. Number of clusters formed in the bubble. }
  \label{figdistrib}%
\end{figure}

A last useful information is the evolution of the cluster size
distribution during the supersaturation pulse. The size distribution
non-dimensionalized by the equilibrium distribution
$ \csd(\gclust, t) / \csdeq(\gclust)$ calculated from
Eq.~(\ref{yshi}), is displayed on Fig.~\ref{figdistrib}a, for
metastable water at $\tlneg$ in the same conditions as
Fig.~\ref{figS}, for 10 equidistant times covering the supersaturation
pulse (thin lines). The classical steady-state $\erfc$ function is
also displayed in thick solid line and the critical size is
materialized by the vertical dashed line. It can be clearly seen that
the steady-state is not reached, which demonstrates that the latter
assumption would be difficult to justify in the present problem. 

Another interesting result is the order of magnitude of the largest
cluster formed in the bubble during the supersaturation pulse. The
number of clusters of each size formed in the bubble was calculated by:
\begin{equation*}
  \nbclust (\gclust, t) = \csd (\gclust, t) V_m
\end{equation*}
where $\csd$ was estimated from Eq.~(\ref{yshi}), and $V_m$ is the
mean volume of the bubble during the equivalent supersaturation
pulse. The result is displayed as thin lines on Fig.~\ref{figdistrib}b
(time increasing from bottom to top). The largest cluster formed is
seen to reach only a dozen of monomers.

%
\section{Discussion}
%

Using a classical reduced model of bubble dynamics including heat and
mass transfer, we have shown that in typical inertial cavitation
conditions at 20~kHz, the water vapor in an air bubble becomes
strongly metastable during the bubble second rebound, both against
liquefaction and freezing.  Approximate nucleation calculations show
that vapor-liquid phase transition indeed takes place and that a few
percents of the initial water content clusterize during the
supersaturation pulse. However, owing to the huge supersaturation
level, the critical size is about 5 molecules and the largest cluster
formed reaches roughly 10 molecules. It is therefore difficult to
conclude to the formation of real droplets, in contrast with the
experiments in supersonic expansion, where clusters of several
thousands molecules are formed, and have additional time to freeze
homogeneously. This is due to the typical time scales involved in the
respective experiments: the full process in nozzles lasts for about
tens of microseconds, whereas the metastable state of the vapor in
the bubble lingers for less than 1~$\mu$s (see caption of
Fig.~\ref{figphaseTambTneg}). Since nucleation is a kinetic process,
it cannot proceed efficiently over such a small timescale.


Homogeneous freezing of such small liquid clusters is therefore very
unlikely and the scheme of ice crystals nucleating \emph{inside} the
bubble, as was initially suggested by Fig.~\ref{figphaseTambTneg} must
be abandoned. However, the fate of the small clusters formed remains
an interesting issue. As supersaturation decreases after the peak, one
may expect these clusters to dissociate back to monomers (after about
600 ns in the case of Fig.~\ref{figS}). This short lifetime casts some
doubts on their ability to redisperse in the surrounding liquid.  We
note however that the same issue can be raised about the redispersion
of the radicals produced by water sonolysis during the bubble
collapse, in the frame of sonochemistry. Indeed, such radicals are
also short-lived species whose precise mechanism of redispersion in
the liquid after the bubble collapse remains unclear. In spite of this
lack of knowledge, the entrance of OH$^\circ$ radicals in the liquid
is the mechanism commonly put forward to explain their contribution to
chemical reactions in the liquid phase.
Whether the liquid clusters formed in the bubble are actually able to
hit the surrounding liquid pertains to the same questioning. If for
the purpose of reasoning, one does not exclude such a mechanism, a
highly supercooled liquid cluster coming into contact with the
surrounding supercooled liquid might act as a pre-existing entity able
to trigger freezing. All would happen as if some part of the
surrounding supercooled liquid (typically at $\tlneg$) underwent a
sudden fluctuation down to minus tens of Celsius. Among the plausible
contacting mechanisms, non-symmetrical bubble collapse or bubble
coalescence may be invoked. Of course, we do not claim that all
bubbles undergo such an event precisely when clusters have formed
inside, but that such an event has a nonzero probability over billions
of bubbles collapsing 20000 times per second. The whole process
constitutes an alternative explanation of the ability of cavitation to
trigger freezing in supercooled liquids.

Whatever the existence of such a process, the large cooling of the
gas/vapor mixture during the bubble rebounds constitutes an
interesting and, as far as the authors are aware, unreported feature
of acoustic cavitation. More complete models of the bubble interior
based on direct Navier-Stokes simulations \cite{storeyszeri2000} would
be required to confirm and quantify more precisely the reached level
of supercooling. If the answer is affirmative, the cavitation bubble
would (again) constitute a rather uncommon physical object, whose
content is able to cool down to temperatures seldom encountered for
water vapor, after having heated up to thousands of Kelvins.

\appendix

%
\section{Calculation of the vapor spinodal curve}
\label{annSpinodale}

The vapor spinodal curve is generally poorly documented in
thermodynamic database. We use therefore the approach of Kim and
co-workers \cite{KimWyslouzil2004} who assumed that the ratio
$\pspin(T/T_C)/\psat(T/T_C)$ for the fluid of interest was the same as
that obtained for an equation of state (EOS) describing a hard sphere
fluid corrected by an attractive Yukawa potential
\cite{LiWilemski2003}. We recall the main lines of calculations in
this appendix. 

\begin{figure}[h!tb]
  \includegraphics[width=0.9\linewidth]{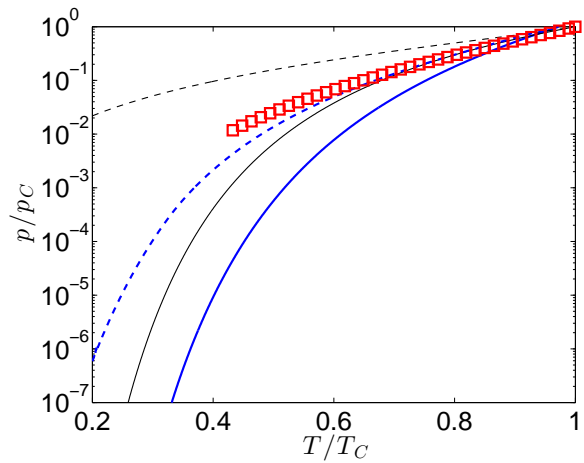}
  \caption{(Color online) Thin (black) lines: vapor pressure curve
    (solid) and vapor spinodal curve (dashed) calculated from hard
    sphere/Yukawa EOS (\ref{pyukawa})-(\ref{muyukawa}). Thick (blue)
    solid line: vapor pressure of water used in this paper
    \cite{MurphyKoop2005}. Thick (blue) dashed line: corrected vapor
    spinodal estimated from (\ref{pspincorr}). The (red) square
    symbols are the vapor spinodal data for water tabulated in
    Ref.~\cite{DobbinsMohammed88}}
    \label{figspinodale}%
\end{figure}

The fluid pressure and chemical potential are given respectively by
the EOS:
\begin{eqnarray}
  \label{pyukawa}
  p (\rho) &=& \rho k T 
  \frac{1 + \eta + \eta^2 - \eta^3}{\left(1 - \eta\right)^3}
  - \frac{1}{2} \alpha \rho^2, \\
  \label{muyukawa}
  \mu_( \rho) &=& k T\left[
    \ln \eta + 
    \frac{8\eta - 9\eta^2 + 3\eta^3}{\left( 1-\eta\right)^3} 
  \right] - \alpha \rho,
\end{eqnarray}
where $\rho$ is the molecular density, $\eta = \pi\sigma^3\rho / 6$ is
the packing fraction, $\sigma$ is the hard sphere diameter, and
$\alpha$ is the amplitude of the attractive potential. The
corresponding dimensionless critical density $\rho_C$ and critical
temperature $T_C$ are found to be:
\begin{equation*}
\rho_C\sigma^3 = 6/\pi\eta_C = 0.24913 
\quad \text{and} \quad
  T_C = \frac{\alpha}{11.1016 \;k\sigma^3 }.
\end{equation*}
The binodal curve is calculated by solving simultaneously
\begin{eqnarray*}
  \mu (T, \rholsat) &=& \mu(T, \rhovsat), \\
  p (T, \rholsat) &=& p(T, \rhovsat), 
\end{eqnarray*}
for the equilibrium densities $\rholsat$ and $\rhovsat$, and the
liquid and vapor spinodal $\rholspin$ and $\rhovspin$ densities are
the roots~of:
\begin{equation*}
  \left(\dsurd{\mu}{\rho} \right)_T = 0.
\end{equation*}
The equilibrium curve and the two spinodal curves in the $(p,T)$ plane
are then obtained by applying the EOS ~(\ref{pyukawa}) to the values
obtained. The equilibrium and vapor spinodal curves obtained are
displayed as thick lines on Fig.~\ref{figspinodale}, and the H$_2$O
equilibrium curve from Ref.~\cite{MurphyKoop2005} used in this paper
in thin solid line. We then assume the real vapor spinodal curve to
be:
\begin{equation}
  \label{pspincorr}
  \pspin = \psat \frac{\pspinyukawa}{\psatyukawa}
\end{equation}
where $\psat$ is the vapor pressure calculated from Ref.~\cite{MurphyKoop2005}.
Figure~\ref{figspinodale} displays the predicted binodal (thin solid
line) and spinodal (thin dashed line) of the hard-sphere Yukawa
equation of state. The binodal of Ref.~\cite{MurphyKoop2005} used
throughout this paper is represented by a thick solid line and the
spinodal deduced from Eq.~(\ref{pspincorr}) by a thick dashed
line. The latter results displays reasonable agreement with the
theoretical prediction of Ref.~\cite{DobbinsMohammed88} (square
symbols).

\vfill
\bibliographystyle{apsrev}



\end{document}